# A Comparison of the Galactic Cosmic Ray Proton and Helium Nuclei Spectra From ~10 MeV/nuc to 1 TeV/nuc Using New Voyager and Higher Energy Magnetic Spectrometer Measurements - Are There Differences In the Source Spectra of These Two Nuclei?

W.R. Webber

New Mexico State University Department of Astronomy, Las Cruces, New Mexico, 88003, USA

# ABSTRACT

This paper determines the relative source spectra of cosmic ray H and He nuclei using a Leaky Box model for galactic propagation and the observed spectra of these nuclei from ~10 MeV/nuc to ~1 TeV/nuc. The observations consist of Voyager 1 measurements up to several hundred MeV/nuc in local interstellar space and measurements above ~10 GeV/nuc where solar modulation effects are small by experiments on BESS, PAMELA and AMS-2. Using BESS and PAMELA measurements which agree with each other, the observed spectra for H and He nuclei and the H/He ratio are well fit by source rigidity spectra for both nuclei which are $\sim P^{-2.24}$ over the entire range of rigidities corresponding to energies between 10 MeV/nuc and several hundred GeV/nuc. In this case, the H/He rigidity source ratio is $5.0 \pm 1$. The recent and presumably more accurate measurements of these spectra above 10 GeV/nuc made by AMS-2 do not entirely agree with the earlier measurements, however. In particular the H spectrum is found to be steeper than that of He by about 0.10 in the spectral exponent. Using the same model for galactic propagation the AMS-2 data leads to source spectra of H and He which are $\sim P^{-2.24}$ up to a break rigidity ~6-8 GV. At higher rigidities the He source spectrum continues to be $\sim P^{-2.24}$ but the required source spectrum for H steepens to an index $\sim P^{-2.36}$ above ~8 GV and, as a result, the H/He source ratio decreases with increasing rigidity using the AMS-2 data.

**<u>Introduction</u>**

The galactic proton and helium components of cosmic rays are the most abundant and perhaps most studied of all of the cosmic ray nuclei. This includes the intensities and relative spectra of these two components. These studies of the comparative spectra have been limited near the Earth, however, by the effects of solar modulation which becomes serious below ~1 GeV/nuc and also by the limited accuracy of the data above 10-20 GeV/nuc where the modulation becomes small. Recent measurements by BESS (Sanuki, et al., 2000) and by the PAMELA spacecraft (Adriani, et al., 2012 ) from ~1.0 GeV/nuc to above ~100 GeV/nuc have provided what appears to be a clear indication of the relative spectra of H and He nuclei above ~10 GeV/nuc to within a few percent. The H/He (E) intensity ratio per energy/nuc from these measurements is shown in Figure 1. Above ~10 GeV/nuc the H/He (E) intensity ratio is ~17.5, more or less constant with energy at higher energies. The decrease in the ratio below ~10 GeV/nuc is due to solar modulation and indicates the severity of these modulation effects which depend mainly on the rigidity of each particle which is different for protons and helium nuclei with the same energy/nuc. Numerous attempts have been made to reconstruct the interstellar spectra of these two nuclei down to lower energies.

A major step in understanding the individual spectra and the intensity ratio at lower energies occurred when V1 crossed the heliopause (HP) in August, 2012. The V1 proton and helium spectra have been described in detail in Cummings, et al., 2016. It was soon realized that the interstellar H/He (E) ratio per E/nuc was essentially constant with energy at a value of about 12 between a few MeV/nuc up to about 350 MeV/nuc (Stone, et al., 2013). This data at lower energies is also shown in Figure 1.

This new Voyager data effectively eliminates the uncertainty due to the solar modulation at the lower energies. But now there is the question of how and why this H/He (E) ratio changes from a nearly constant value ~17.5 above ~10 GeV/nuc to a different relatively constant value ~12.0 below ~350 MeV/nuc. Recall that those ratios are in terms of energy/nuc. If the original "source" spectra of these two nuclei were rigidity spectra then it would be necessary to transform these rigidity spectra into energy/nuc spectra. The transformation between a differential energy intensity and a differential rigidity intensity is given by

$$\left(\frac{dj}{dE}\right) = \frac{1}{2\beta}\left(\frac{dj}{dP}\right)$$

where β is the velocity of the H or He nucleus at the given rigidity P.

This energy dependent transformation is a constant above ~10 GeV/nuc where β→1.0. If, as an example, we start with the measured ratio of 17.5 above ~10 GeV/nuc and multiply it by ½β, then the changing H/He (E) nuclei ratio as a function of energy which corresponds to a constant ratio H/He (P), is given by the dashed line in Figure 1. The good connection between high and low energy measurements using this very simple example suggests that the source spectra of H and He nuclei may have similar rigidity spectra for each component, so that when plotted as energy/nuc spectra, give the measured H/He (E) ratios shown in Figure 1. This would imply that the "source" spectra are, in fact, very similar rigidity spectra for both H and He for the range of energies considered here, with perhaps propagation features dominating below a few hundred MeV/nuc and contributing to the production of a nearly constant H/He (E) nuclei ratio at low energies. In this scenario the H/He (P) ratio of the source spectra could be nearly independent of rigidity with the same spectral exponent for each nucleus and a constant value for the H/He (P) ratio.

Then in 2015, the AMS-2 studies (AMS at CERN, 2015, Aguilar, et al.., 2016a, b) appeared and the high energy part of this picture changed drastically. The AMS-2 value for the H/He ratio per E/nuc was about 16.0 at 10 GeV/nuc (roughly 10% lower than BESS and PAMELA) and then this ratio decreased to a value ~11 at 1 TeV/nuc. For the corresponding H/He (P) ratios the measured values by AMS-2 were ~5.5 at 10 GV decreasing to ~3.5 at 1 TV. This implies that the H and He nuclei rigidity spectra are different above ~10 GV and have spectral exponents that differ by ~0.08-0.10 in this rigidity range. The AMS-2 data is shown in Figure 2.

Figure 2 is identical to Figure 1, showing the Voyager data at lower energies and the AMS-2 data as a function of energy/nuc at higher energies instead of the earlier BESS and PAMELA data. Figures 1 and 2 may be overlaid to see the differences in the higher energy data points. They are:

(1) The absolute value of the H/He (E) ratio from AMS-2 is below that of the BESS and PAMELA measurements at ~10 GeV/nuc by 10-15%. A fraction of this difference could be related to solar modulation effects, which are evident in the larger solar modulation that is observed by AMS-2 at lower energies. This additional solar modulation results in a lower H/He (E) ratio at 1 GeV/nuc for the AMS-2 data, as compared with the BESS and PAMELA data at this energy.

(2) The H/He (E) ratio measured by AMS-2 above ~10 GeV/nuc decreases in a manner consistent with a difference ~0.10 in the exponents of the H and He nuclei spectra. This difference in the exponent is also noted in the individual rigidity spectra of H and He nuclei above ~10 GV presented in AMS at CERN, 2015, (also Aguilar, et al., 2016 a, b). For the BESS and PAMELA data the relatively constant H/He (E) ratio above ~10 GeV/nuc indicates similar E/nuc or rigidity spectra for H and He nuclei at these higher energies.

So the underlying goal of this paper is to "connect" and interpret the H/He (E) ratio measurements at lower energies at Voyager and those above ~10 GeV/nuc made by BESS and PAMELA and also by AMS-2, recognizing that, in the case of the AMS-2 measurements at higher energies, the spectra of the two components should differ by ~0.10 in the exponent and then to reconcile these differing LIS H/He (E) ratios as a function of energy with corresponding LIS H/He (P) ratios as a function of rigidity which are, in fact, the source ratios.

**Interstellar Propagation in a Leaky Box Model – Fitting the BESS and PAMELA and the AMS-2 Data**

We have previously made extensive calculations of the relative proton and helium nuclei spectra to be expected using a Leaky Box propagation model (LBM) for these particles in the galaxy (Webber and Higbie, 2009, 2013; Webber, 2017). In these calculations it has been assumed that all nuclei components have identical rigidity source spectra of the form

$$\left(dj/dP\right) \sim P^{-2.24}$$

For the interstellar propagation of these basic source spectra in a LBM the most recent calculations have been made by comparing the He and C spectra (Webber, 2017). Using the above source spectra for He and C nuclei, the propagation calculations have determined an interstellar mean path length, $\lambda$, which is

$$\lambda = 20.6\, \beta\, P^{-0.45}$$

above a $P_0 \sim 1$ GV. This exponent for the rigidity dependence is based on fitting the AMS-2 B/C ratio measurements in a LBM propagation from ~1 GeV/nuc up to ~1 TeV/nuc (Webber and Villa, 2016).

Below ~1 GV several possibilities have been considered for the rigidity dependence of $\lambda$. The latest (Webber, et al., 2018) is that the mean path length becomes constant or only weakly dependent on P below ~1.0 GV. This assumption is based on the relatively large fraction of purely secondary nuclei such as $^2$H, $^3$He and B that are observed below ~100 MeV/nuc.

In Figure 1 we have superimposed (in blue) the LBM calculation of the propagated LIS H/He (E) ratio along with the ratio measured by Voyager at low energies and by BESS and PAMELA at higher energies. This LBM calculation is based on identical source rigidity spectra ~$P^{-2.24}$ for H and He nuclei at all rigidities and with the path length $\lambda=9.0$ g/cm$^2$ below $P_0=1.0$ GV. The calculated ratio is fit to the measured value of H/He (E)=12.5 by Voyager at ~100 MeV/nuc. This calculation predicts a H/He (E) ratio that increases from ~12.5 to a value ~17.5 above 10 GeV/nuc and remains nearly constant above 10 GeV/nuc and is therefore consistent with the BESS and PAMELA data at higher energies. The source ratio, H/He (P), is $5.0 \pm 1$ at all rigidities in this calculation.

Next we fit the decreasing H/He (E) ratio above ~10 GeV/nuc that is measured by AMS-2. These results of these calculations are shown in Figure 2 for a source proton rigidity spectrum which is now taken to be ~$P^{-2.36}$ above 10 GV, 0.12 power steeper than that for He which remains ~$P^{-2.24}$ at all rigidities. It is evident from these H/He (E) ratio calculations that a "source" spectrum for protons that is similar to ~$P^{-2.24}$ at lower rigidities changing to one ~$P^{-2.36}$ at higher rigidities is necessary to simultaneously fit both the new AMS-2 data and that from Voyager.

To identify the rigidity at which a simple break in the interstellar proton spectra occurs and to place limits on the spectral index of the low rigidity part of the proton spectrum that will match Voyager data and the high energy AMS-2 data, we next assume proton source spectral indices in rigidity between -2.20 and -2.28 below 10 GeV/nuc along with the same value of $P_0$=1.0 for the change in the rigidity dependence of the path length. Examples of the H/He (E) ratio obtained from these calculations for source spectral indices = 2.20, 2.24 and 2.28 at lower energies are shown in Figure 3. Normalizing the H/He (E) ratio obtained from these calculations to the Voyager measured value of 12.5 in each case for the ratio at 100 MeV/nuc, leads to a ratio 19.4 at 10 GeV/nuc for a proton spectrum with index -2.20 below ~10 GeV/nuc and 15.3 for a proton spectrum with index -2.28 below ~10 GeV/nuc. The AMS-2 measurement at 10 GeV/nuc is, in fact, 15.3 and this measured ratio includes significant effects due to solar modulation. So to further understand the details of the spectral break we now need also to consider the effects of solar modulation.

**<u>The Effects of Solar Modulation on the Spectra and Ratios</u>**

To calculate the effects of solar modulation we use the Gleeson and Axford, 1968, approximation to describe this modulation (see e.g., Usoskin, et al., 2014). In this calculation the intensity j (E) at the Earth, or any location in the heliosphere (Webber, et al., 2017), is given by

$$j(E) = j(E + \phi) \cdot [E(E + 2E_0)/(E + \phi)(E + \phi + 2E_0)]$$

where $\phi$ in MV equals the modulation potential between local IS space and the location where the measurement is made and $E_0$ is the proton rest energy = 938 MeV.

In Figure 4 we show the effects of this solar modulation on the H/He (E) ratio, where the values of $\phi$ =100, 400 and 660 MV for the modulation potential are used, along with a source spectral index ~-2.24 in rigidity.

Note that for a solar modulation $\phi$=400 MV, the calculated H/He (E) ratios agree well with those measured by BESS and PAMELA below ~10 GeV/nuc (shown in Figure 1) at times when the overall modulation is at a minimum at the Earth. These modulation calculations also show that at 10 GeV/nuc a modulation potential of 400 MV changes the H/He (E) ratio from its calculated LIS value of 17.7 for a LBM propagation with $\lambda$=20.6 $\beta P^{-0.45}$ > 1.0 GV and $\lambda$=9g/cm$^2$ constant below 1.0 GV, to a value of 15.6 (a decrease of ~11%).

The LIS values for the H/He (E) ratio from this particular calculation at lower energies also agree well with recent V1 measurements of this ratio, which include new data up to ~650 MeV/nuc, for both protons and helium nuclei from V1. The (calculated) LIS H/He (E) ratio is seen to increase suddenly from ~12 to ~15 at just above 400 MeV/nuc. This is due to the choice of $P_0$=1.0 GV. This is just the rigidity of a 440 MeV proton, so the lower energy protons begin to respond to the changed values of the diffusion coefficient below 1.0 GV, whereas the He nuclei do not respond until their energy becomes less than ~125 MeV/nuc at which point their rigidity is 1.0 GV.

The two highest energy proton measurements at ~420 and 570 MeV/nuc for V1 shown in Figure 4 were not included in the original presentation of the H/He (E) ratio measured at V1 (Stone, et al., 2013). If they are included they show a sudden jump up of the measured LIS H/He (E) ratio above ~400 MeV/nuc, giving support to the assumption that there is indeed a break in the interstellar diffusion coefficient at ~1.0 GV or slightly below.

The LIS H and He intensities per E/nuc calculated using a LBM from this new study, along with the resulting H/He (E) ratio are listed in Table 1. Also included in Table 1 are the LIS spectra of $^2$H and $^3$He secondaries. Note that the intensity in E/nuc of $^3$He becomes ~20% of that of $^4$He at energies between 1-10 GeV/nuc and therefore the secondary contribution to $^4$He is important (see Webber, et al., 2018, for a determination of the $^2$H and $^3$He intensities at Voyager and their interpretation using the same set of LBM propagation calculations).

In Figure 5 we show both the H and He LIS rigidity spectra derived from the corresponding energy spectra shown in Table 1. These spectra have a relatively constant difference in intensity of a factor of between 4-6 above a few GV, possibly decreasing above 10 GV due to a steeper proton spectrum. Below ~1 GV the H and He rigidity spectra show a large difference due to the conversion between energy and rigidity and also because of ionization energy loss during propagation.

**The H and He Nuclei Source Rigidity Spectra**

Regarding the source spectra of H and He nuclei from the above discussion, there are really two scenarios. The simplest scenario that fits both the data from Voyager and that from BESS and PAMELA is one in which the proton and Helium nuclei both have the same source rigidity spectrum, $dj/dp \sim P^{-2.24}$. The ratio of source intensities H/He (P) is $5.0 \pm 1$.

The second scenario requires that, at higher rigidities, the proton source spectrum becomes $\sim P^{-2.36}$ but the He source spectral exponent is unchanged, thus leading to a gradually decreasing H/He (P) ratio at higher rigidities.

Considering these alternatives we conclude then that: (1) If one matches the measured Voyager, BESS and PAMELA H/He (E) ratios and the calculated H/He (E) ratios as a function of energy, after interstellar propagation in a LBM with $\lambda = 20.6 \beta P^{-0.45}$ and solar modulation ~400 MV, then a proton source spectrum $\sim P^{-2.24}$ along with a helium spectrum also $\sim P^{-2.24}$ with an intensity ratio, H/He (P) $5.0 \pm 1$ for the source spectra, independent of rigidity, are required to fit the data.

(2) In the case of the AMS-2 measured H/He ratios $\geq 10$ GV, because of the subtle interplay between the source spectral differences between H and He nuclei above ~10 GV measured by AMS-2 and the effects of interstellar propagation and the solar modulation, and using the Voyager data at lower energies, the location in energy (rigidity) of the spectral break required to fit the proton spectrum is between 6-8 GV along with a spectral difference between H and He nuclei of $0.10 \pm 0.02$ above ~8 GV. The H/He (P) source ratio in this scenario thus decreases from the AMS-2 measured value ~5.5 at 10 GV to a value ~3.5 at 1 TV because of the different source spectral indices of H and He nuclei at higher rigidities.

In these scenarios, on a scale of total nucleons below 10 GV, a H/He (P) source ratio of 5.0 would imply that 56% of the accelerated nucleons are H and 44% are He nucleons. At 1 TV these fractions would be 48% H and 52% He. These fractions may be compared with the typical cosmological fractions obtained in the Big Bang scenario of 76% H and 24% He, for example.

Are these H and He source abundance ratios that we determine, which are changing with energy (rigidity) above 6-8 GV, due to a selection process in the acceleration itself or are they simply a result of the local "source" abundances themselves which change, in the AMS alternative, for example, as the "source shock" moves outwards through different abundance regions in the "supernova"?

And finally, why does the He spectrum and also the C spectrum, both of which are A/Z=2 nuclei, show no evidence of a spectral break (discussed in Webber, 2017), whereas the protons and also the electrons which are discussed earlier in Webber and Villa, 2017, both show evidence of a possible spectral break of about the same magnitude of 0.10 in the exponent and also at about the same value of rigidity, ~6-8 GV when the AMS-2 measurements of these components are used?

Figure 6 shows the LIS H rigidity spectrum (also shown in Figure 5) along with the H source rigidity spectrum. The shaded region highlighting the difference between them is entirely due to propagation effects in the galaxy. These include, at low rigidities, the ionization energy loss effect and at higher rigidities the diffusion loss from the galaxy, which is $\sim P^{0.45}$, as well as nuclear interactions.

## Summary and Conclusions

New measurements of the proton and helium nuclei spectra between ~10 GeV/nuc to 1 TeV/nuc by AMS-2 have indicated that the spectrum of cosmic ray protons is about 0.10 power steeper in the spectral index than that of He nuclei. This is in contrast to earlier measurements made by the PAMELA and BESS instruments which show similar spectra for H and He nuclei above 10 GeV/nuc and which are also consistent with Voyager measurements of the P/He (E) ratio below 300 MeV/nuc; assuming that the source spectra of both P and He are rigidity spectra which are $\sim P^{-2.24}$, independent of rigidity.

In the case of the AMS-2 measurements, starting with the premise that the source spectra of both components are given by rigidity spectra of the form dj/dP $\sim P^{-2.24}$ below ~10 GV, we have shown that the P/He (E) ratio as a function of energy observed by AMS-2 above 10 GeV/nuc and the ratio measured by Voyager at low energies may be explained if the source spectral exponent of He is independent of rigidity at all rigidities whereas the exponent for proton source spectrum changes from -2.24 at lower rigidities to -2.36 at higher rigidities.

This scenario assumes propagation in a LBM where the path length $\lambda = 20.6 \beta P^{-0.45}$ above 1.0 GV. Below 1 GV this path length becomes a constant ~9.0 g/cm$^2$. The effects of solar modulation on this P/He (E) ratio are considered and allow us to accurately predict, using a modulation potential ~660 MV, the P/He (E) ratio that is measured by AMS-2 down to ~1 GeV/nuc.

In either scenario the H/He (P) source abundance ratio is $5.0 \pm 1$ below the ~6-8 GV. In the AMS-2 scenario this source abundance ratio above 10 GV will decrease to ~3.5 at 1 TV.

The break in the path length rigidity dependence at 1 GV assumed in the above calculations is a result of a change in the rigidity dependence of the interstellar diffusion coefficient at this rigidity. This causes the calculated LIS P/He (E) ratio to increase suddenly from ~12 to 15 at about 400 MeV/nuc, which is, in fact, the rigidity of a 1 GV proton. We have used recent Voyager H/He (E) measurements up to ~650 MeV/nuc to confirm the existence of this predicted jump in the data.

In a previous paper we have compared the intensities of He and C nuclei measured down to ~10 MeV/nuc at Voyager with those for these nuclei measured by AMS-2 up to ~1 TeV/nuc. We found that, using the same LBM for interstellar propagation described above and identical source spectra $\sim P^{-2.24}$ for both He and C nuclei, we could obtain propagated intensities of both components that agree to within a few % with the measurements at both low and high ends of the spectrum spanning a range of $10^5$ in energy (Webber, 2017), thus implying $P^{-2.24}$ source spectra for both nuclei extending over the entire rigidity range.

The difference in the source proton spectrum relative to that of helium found using AMS-2 data at high rigidities may indicate a pattern in which the source spectra of A/Z ~2.0 cosmic ray nuclei have nearly identical $P^{-2.24}$ rigidity spectra at both lower and higher rigidities, of which He and C are the only examples available for study at this time. The source proton spectra has a break at a rigidity ~6-8 GV if one uses AMS-2 high rigidity data as a base.

If the BESS and PAMELA measurements of the proton and helium spectra above ~10 GV are used as a data base these differences between the source nuclei spectra of protons and heavier cosmic ray nuclei almost completely vanish.

**Acknowledgements:** The authors are grateful to the Voyager team that designed and built the CRS experiment with the hope that one day it would measure the galactic spectra of nuclei and electrons. This includes the present team with Ed Stone as PI, Alan Cummings, Nand Lal and Bryant Heikkila, and to others who are no longer members of the team, F.B. McDonald and R.E. Vogt. Their prescience will not be forgotten. This work has been supported throughout the more than 40 years since the launch of the Voyagers by the JPL.

| TABLE 1 | | | | | | |
|---|---|---|---|---|---|---|
| Energy | H1 | H2 | He3 | He4 | Sum He | H1/Sum He |
| 1.76 | 1.14E+01 | 4.06E-02 | 1.50E-02 | 8.79E-01 | 8.94E-01 | 1.28E+01 |
| 2.37 | 1.40E+01 | 5.12E-02 | 1.98E-02 | 1.07E+00 | 1.09E+00 | 1.28E+01 |
| 3.16 | 1.68E+01 | 6.40E-02 | 2.48E-02 | 1.28E+00 | 1.31E+00 | 1.28E+01 |
| 4.27 | 1.97E+01 | 8.53E-02 | 3.17E-02 | 1.51E+00 | 1.54E+00 | 1.28E+01 |
| 5.6 | 2.29E+01 | 1.05E-01 | 4.00E-02 | 1.75E+00 | 1.79E+00 | 1.28E+01 |
| 7.5 | 2.60E+01 | 1.30E-01 | 5.08E-02 | 1.98E+00 | 2.03E+00 | 1.28E+01 |
| 10 | 2.84E+01 | 1.59E-01 | 6.45E-02 | 2.22E+00 | 2.25E+00 | 1.26E+01 |
| 13 | 3.09E+01 | 1.91E-01 | 8.08E-02 | 2.40E+00 | 2.45E+00 | 1.26E+01 |
| 18 | 3.26E+01 | 2.24E-01 | 9.99E-02 | 2.52E+00 | 2.59E+00 | 1.26E+01 |
| 24 | 3.37E+01 | 2.54E-01 | 1.20E-01 | 2.57E+00 | 2.65E+00 | 1.27E+01 |
| 32 | 3.40E+01 | 2.79E-01 | 1.40E-01 | 2.56E+00 | 2.68E+00 | 1.27E+01 |
| 42 | 3.25E+01 | 2.93E-01 | 1.58E-01 | 2.46E+00 | 2.56E+00 | 1.27E+01 |
| 56 | 3.05E+01 | 2.93E-01 | 1.69E-01 | 2.30E+00 | 2.38E+00 | 1.28E+01 |
| 75 | 2.78E+01 | 2.81E-01 | 1.75E-01 | 2.07E+00 | 2.17E+00 | 1.28E+01 |
| 100 | 2.39E+01 | 2.57E-01 | 1.71E-01 | 1.79E+00 | 1.87E+00 | 1.28E+01 |
| 133 | 2.02E+01 | 2.36E-01 | 1.56E-01 | 1.52E+00 | 1.60E+00 | 1.26E+01 |
| 178 | 1.59E+01 | 2.03E-01 | 1.43E-01 | 1.23E+00 | 1.30E+00 | 1.22E+01 |
| 237 | 1.19E+01 | 1.68E-01 | 1.35E-01 | 9.52E-01 | 1.01E+00 | 1.18E+01 |
| 316 | 8.62E+00 | 1.32E-01 | 1.16E-01 | 7.00E-01 | 7.56E-01 | 1.14E+01 |
| 422 | 5.95E+00 | 9.71E-02 | 8.74E-02 | 4.88E-01 | 5.31E-01 | 1.12E+01 |
| 562 | 5.24E+00 | 6.64E-02 | 6.14E-02 | 3.23E-01 | 3.54E-01 | 1.48E+01 |
| 750 | 3.42E+00 | 4.28E-02 | 4.04E-02 | 2.04E-01 | 2.25E-01 | 1.52E+01 |
| 1000 | 2.14E+00 | 2.65E-02 | 2.52E-02 | 1.25E-01 | 1.37E-01 | 1.56E+01 |
| 1334 | 1.34E+00 | 1.57E-02 | 1.53E-02 | 7.58E-02 | 8.50E-02 | 1.58E+01 |
| 1779 | 7.89E-01 | 9.01E-03 | 8.91E-03 | 4.46E-02 | 4.90E-02 | 1.61E+01 |
| 2372 | 4.47E-01 | 4.95E-03 | 4.99E-03 | 2.55E-02 | 2.74E-02 | 1.63E+01 |
| 3163 | 2.53E-01 | 2.62E-03 | 2.69E-03 | 1.42E-02 | 1.55E-02 | 1.63E+01 |
| 4218 | 1.39E-01 | 1.33E-03 | 1.40E-03 | 7.71E-03 | 8.30E-03 | 1.68E+01 |
| 5624 | 7.44E-02 | 6.58E-04 | 7.03E-04 | 4.08E-03 | 4.35E-03 | 1.71E+01 |
| 7500 | 3.95E-02 | 3.16E-04 | 3.42E-04 | 2.11E-03 | 2.27E-03 | 1.74E+01 |
| 10001 | 2.01E-02 | 1.47E-04 | 1.62E-04 | 1.07E-03 | 1.15E-03 | 1.75E+01 |
| 13337 | 9.98E-03 | 6.70E-05 | 7.45E-05 | 5.33E-04 | 5.70E-04 | 1.75E+01 |
| 17785 | 4.82E-03 | 2.98E-05 | 3.36E-05 | 2.61E-04 | 2.77E-04 | 1.74E+01 |
| 23717 | 2.32E-03 | 1.31E-05 | 1.48E-05 | 1.27E-04 | 1.35E-04 | 1.72E+01 |
| 31627 | 1.08E-03 | 5.63E-06 | 6.45E-06 | 6.08E-05 | 6.40E-05 | 1.68E+01 |
| 42176 | 4.95E-04 | 2.40E-06 | 2.77E-06 | 2.90E-05 | 3.02E-05 | 1.64E+01 |
| 56242 | 2.27E-04 | 1.01E-06 | 1.18E-06 | 1.37E-05 | 1.43E-05 | 1.59E+01 |
| 75000 | 1.03E-04 | 4.25E-07 | 4.98E-07 | 6.45E-06 | 6.72E-06 | 1.54E+01 |
| 100014 | 4.63E-05 | 1.78E-07 | 2.09E-07 | 3.02E-06 | 3.13E-06 | 1.48E+01 |
| 133371 | 2.09E-05 | 7.37E-08 | 8.71E-08 | 1.41E-06 | 1.45E-06 | 1.44E+01 |
| 177853 | 9.44E-06 | 3.05E-08 | 3.62E-08 | 6.56E-07 | 6.74E-07 | 1.40E+01 |
| 237171 | 4.23E-06 | 1.26E-08 | 1.50E-08 | 3.04E-07 | 3.11E-07 | 1.36E+01 |
| 316273 | 1.81E-06 | 5.17E-09 | 6.17E-09 | 1.35E-07 | 1.38E-07 | 1.31E+01 |
| 421757 | 8.14E-07 | 2.12E-09 | 2.54E-09 | 6.23E-08 | 6.36E-08 | 1.28E+01 |
| 562422 | 3.62E-07 | 8.69E-10 | 1.04E-09 | 2.87E-08 | 2.92E-08 | 1.24E+01 |
| 750001 | 1.62E-07 | 3.55E-10 | 4.27E-10 | 1.32E-08 | 1.34E-08 | 1.21E+01 |
| 1000143 | 7.27E-08 | 1.45E-10 | 1.75E-10 | 6.08E-09 | 6.16E-09 | 1.18E+01 |

Intensities are in P/m$^2$·sr·s·MeV/nuc

# FIGURE CAPTIONS

**Figure 1:** The H/He (E) nuclei ratios of intensities measured by BESS (Sanuki, et al., 2000) and PAMELA (Adriani, et al., 2012) at the Earth at a time of minimum solar modulation. Also shown in the figure are the LIS ratios of intensities measured by V1 beyond the HP at lower energies (Stone, et al., 2013). The dashed line shows how source rigidity spectra with the same spectral exponent for both H and He nuclei would be modified into energy/nuc spectra by the factor ½ β where the ratio per E/nuc is normalized to a value 17.5 at 10 GeV/nuc. The solid blue line is a LBM calculation described in the text.

**Figure 2:** Similar to Figure 1 using V1 data for the P/He (E) ratio at low energies, but now using the AMS-2 values of this ratio per E/nuc (AMS at CERN, 2015; Aguilar, et al., 2016a, b) at energies $\geq$ 1 GeV/nuc. Note the differences in high energy H/He (E) ratio measurements with those in Figure 1. These differences are discussed in the text. The solid blue lines show LBM calculations where the proton source spectrum is taken to be $\sim P^{-2.24}$ at low rigidities and $\sim P^{-2.36}$ at high rigidities.

**Figure 3:** Similar to the data shown in Figure 2 except now 3 possible source low rigidity spectra are considered for protons with exponents $\sim -2.20$, $-2.24$ and $-2.28$ below 10 GeV, each normalized to a value of 12.5 at 100 MeV/nuc for the H/He (E) ratio.

**Figure 4:** The effects of solar modulation on the calculated LIS H/He (E) ratio. The propagated LIS ratio with source spectra $\sim P^{-2.24}$ for both nuclei is shown along with values of the ratio calculated using modulation potentials $\phi = 100$, 400 and 660 MV. The V1 data on the LIS H/He (E) ratio is for the overall time period from 2013 to 2017 and includes proton and helium intensities up to 650 MeV/nuc.

**Figure 5:** The LIS rigidity spectra of H and He nuclei. This data is obtained by transforming the energy/nuc calculations in Table 1. Note the peaks in the H and He rigidity spectra below $\sim 1$ GV and separated by a factor $\sim 2$ in rigidity. Above 1 GV the ratio of the two spectra becomes $\sim 5.0$. Above 10 GV this ratio of 5.0 becomes smaller as a result of the steeper H spectrum measured by AMS-2.

**Figure 6:** This figure shows the LIS H rigidity spectrum also shown in Figure 5 and the derived source rigidity spectrum for H which is ~$P^{-2.24}$ below 6-8 GV and ~$P^{-2.36}$ at higher rigidities. The difference between these two spectra, shown as a shaded region, is due to propagation effects in the galaxy ranging from ionization energy loss in the matter disk at lower rigidities to diffusion in and out of the galaxy, defined by a $P^{0.45}$ dependence at rigidities >1 GV changing to a ~$P^{-1.0}$ dependence at the lowest rigidities and also to other loss processes included in the propagation code.

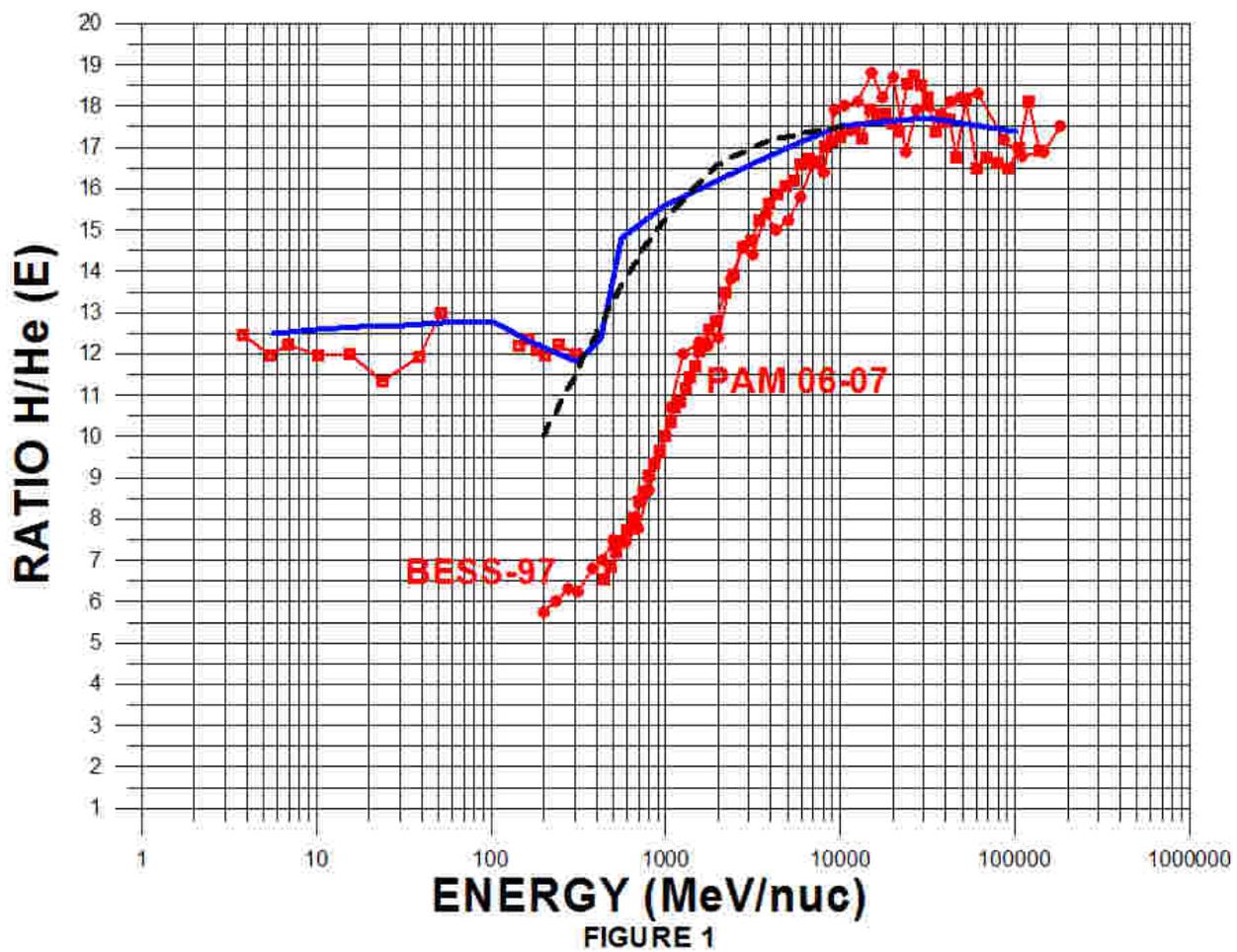

**FIGURE 1**

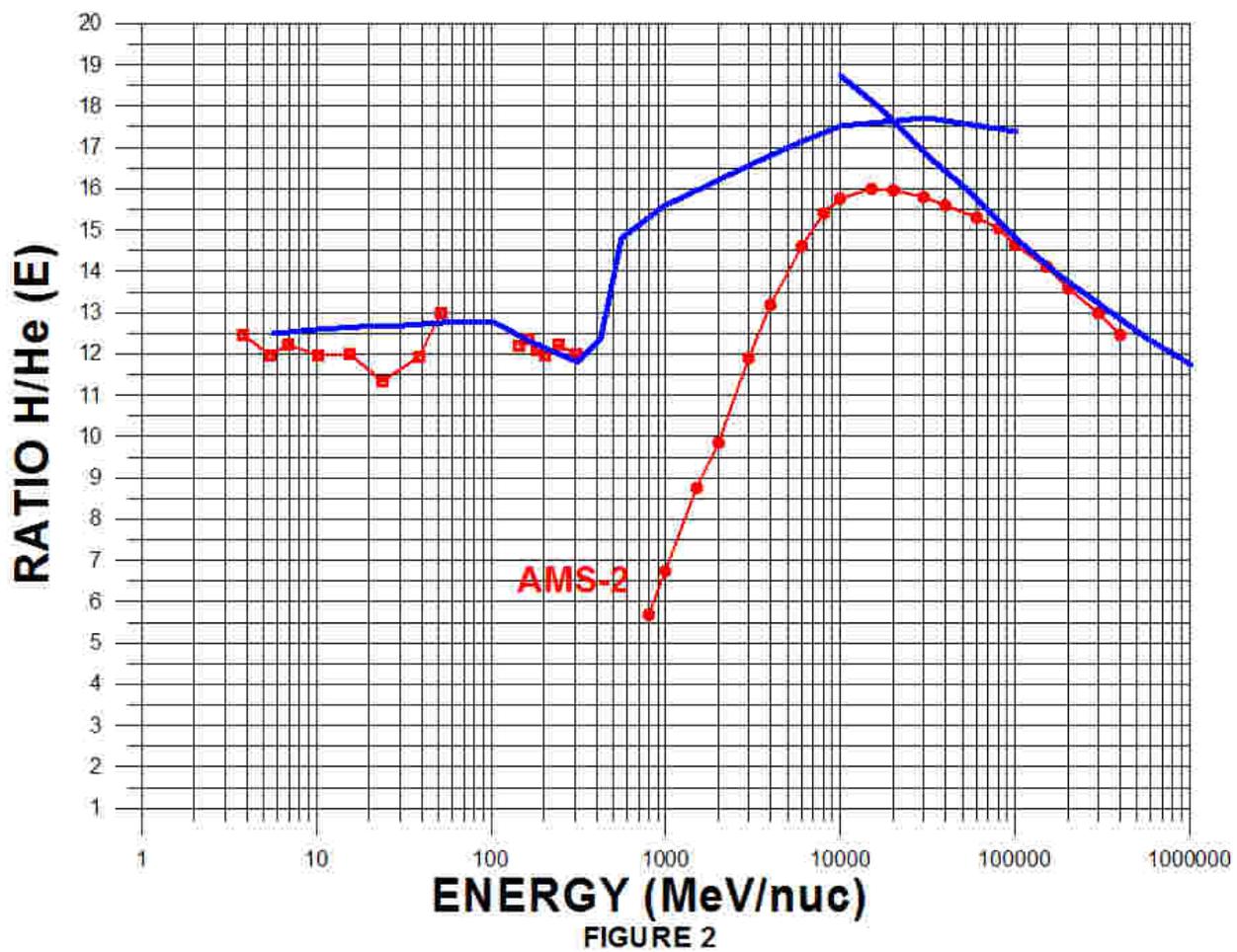

FIGURE 2

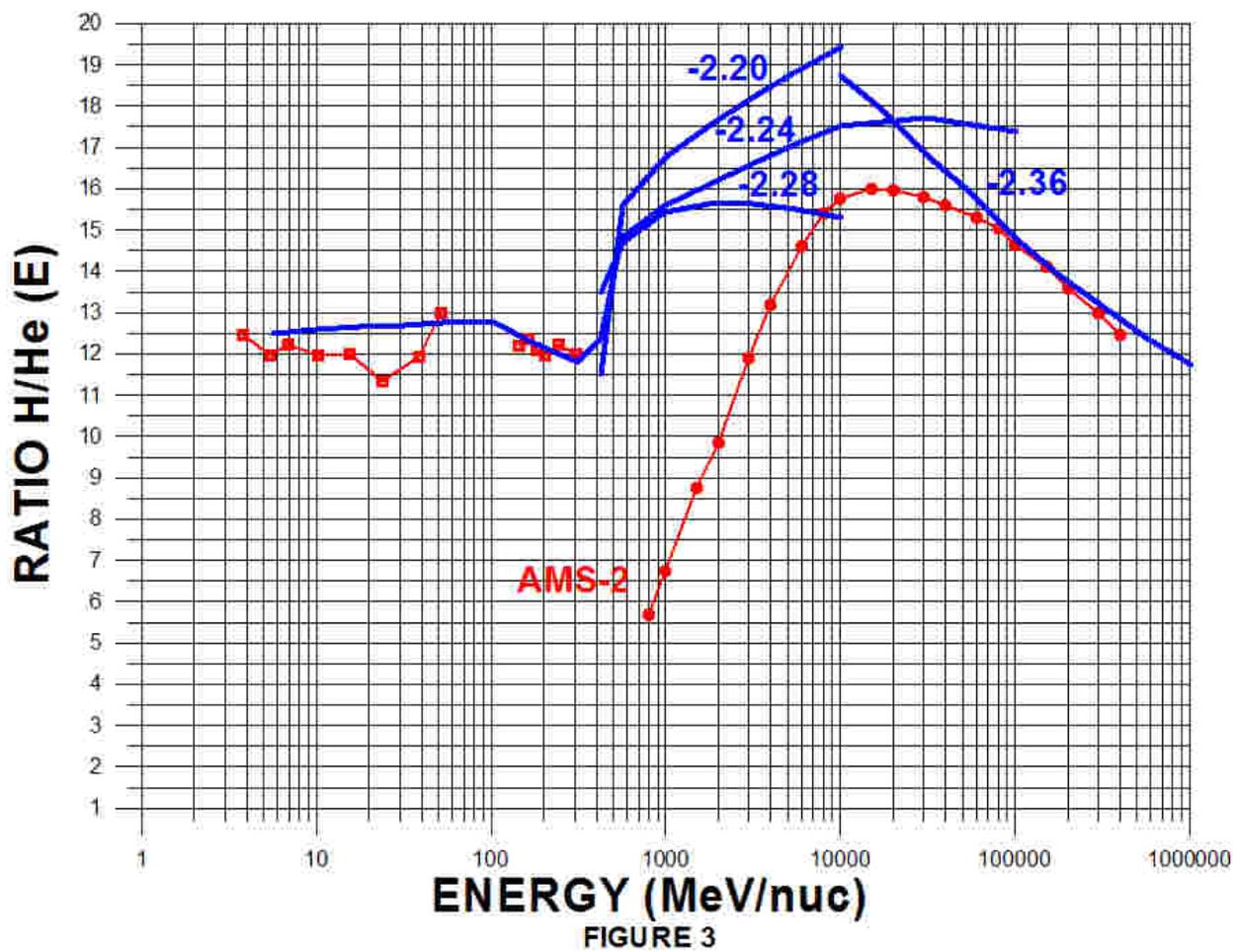

FIGURE 3

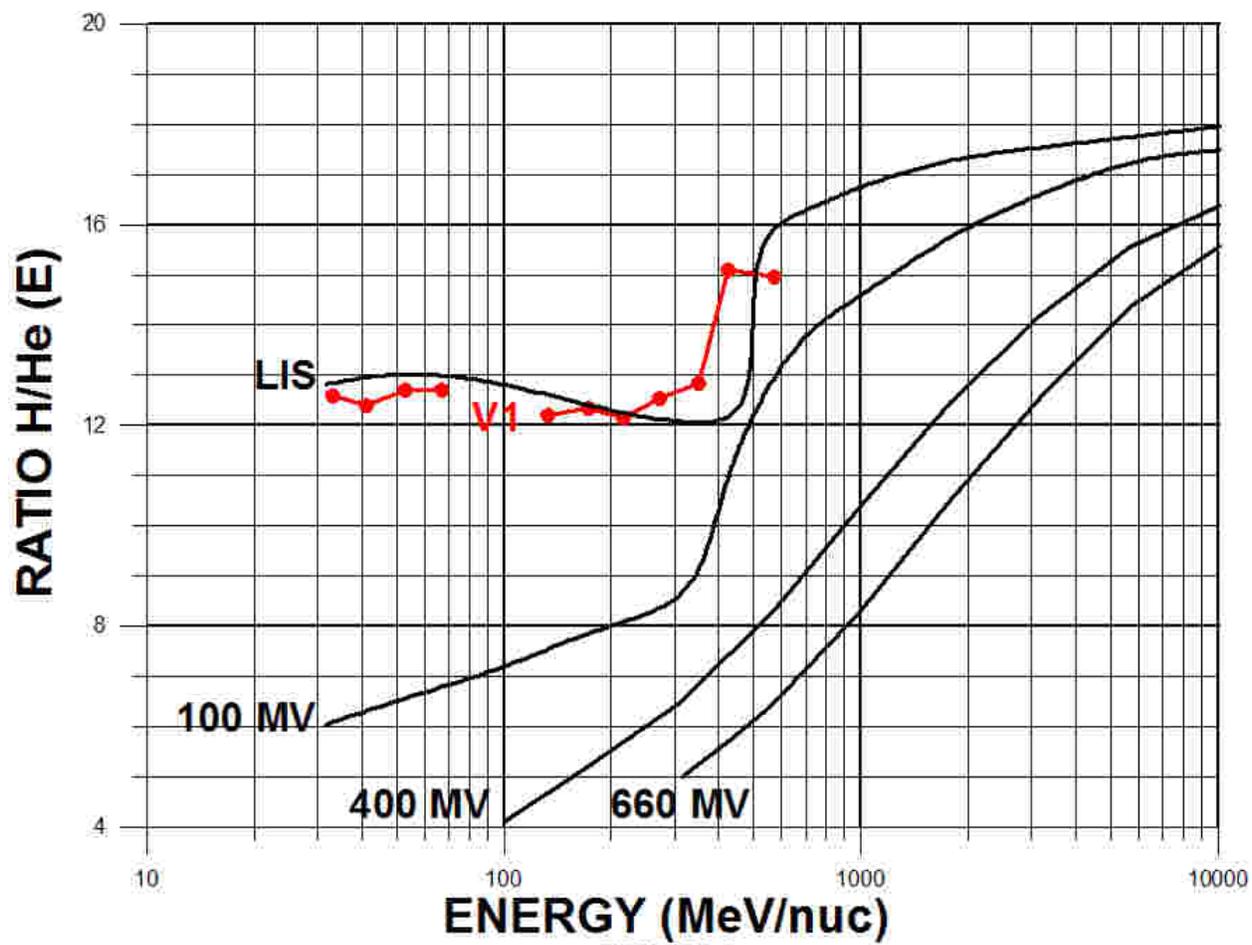

FIGURE 4

<->
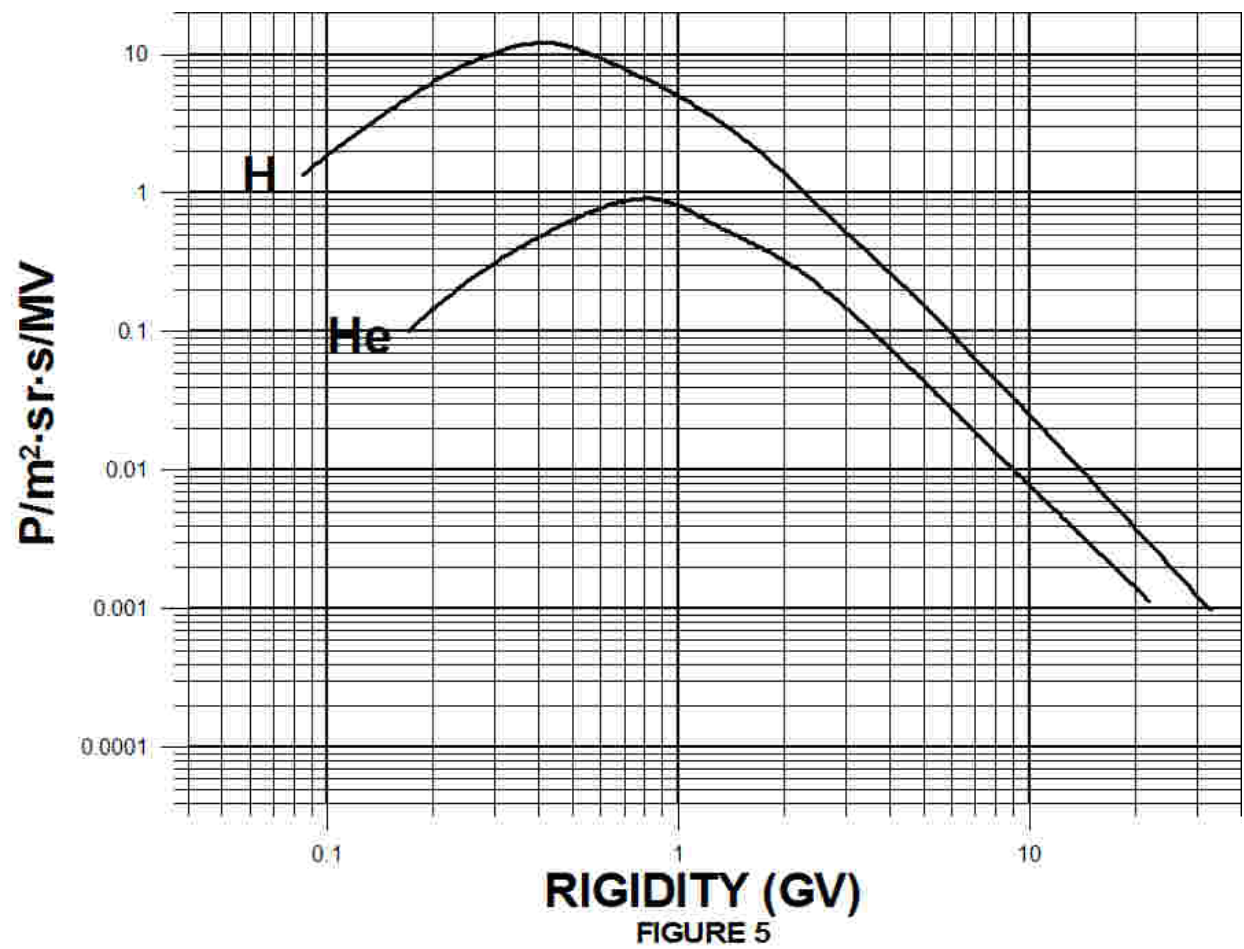

FIGURE 5

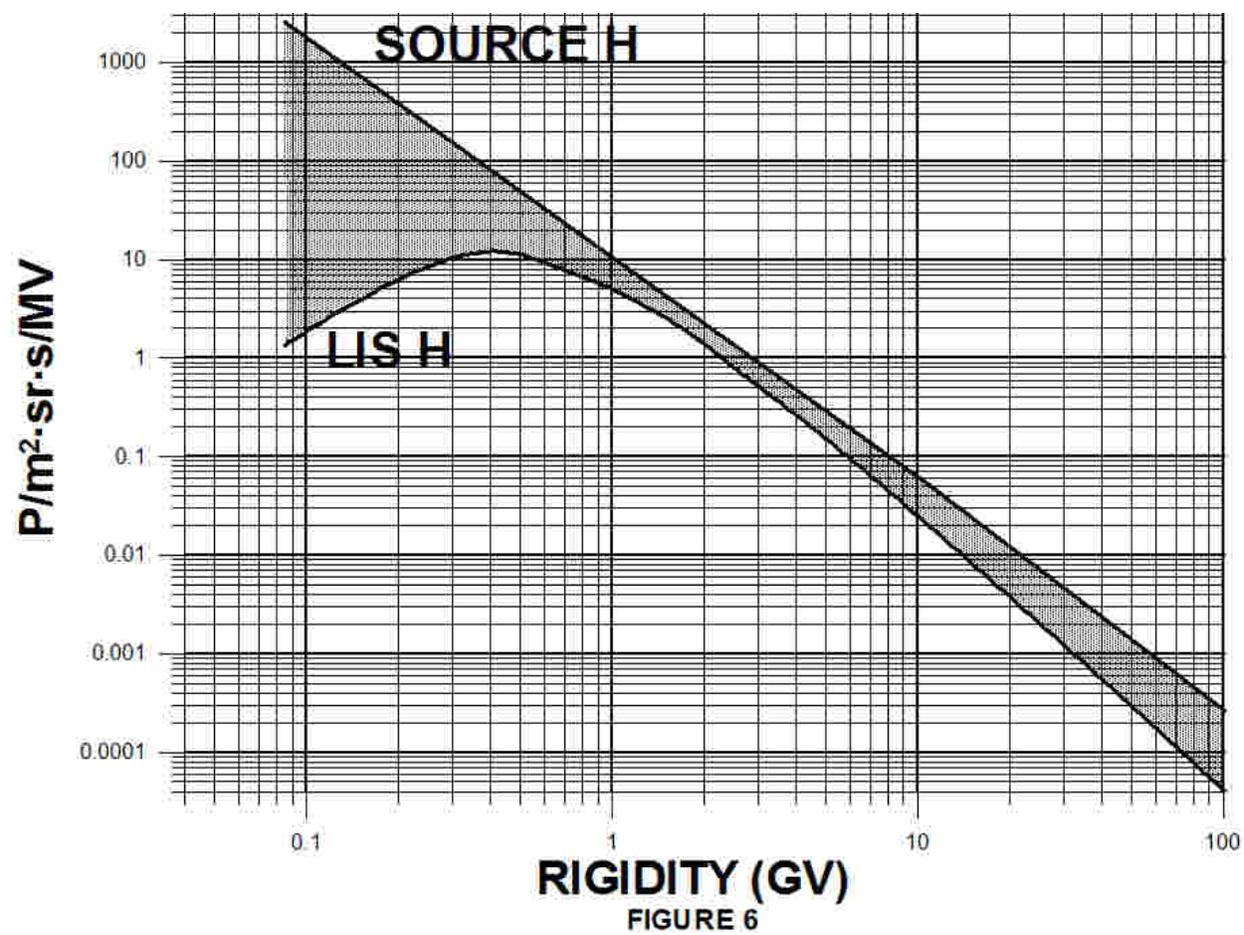

FIGURE 6